# Averaging of Spontaneous Radiation in Relativistic Strophotron

I.V. Dovgan*

Department of Physics, Moscow State Pedagogical University, Moscow 119992, Russia.

*dovganirv@gmail.com

**Abstract.** The main resonance frequency is shown to depend on the initial conditions of the electron, and in particular on its initial transversal coordinate xo. This dependence $\omega_{res}(x_0)$ is shown to give rise to a very strong inhomogeneous broadening of the spectral lines. The broadening can become large enough for the spectral lines to overlap with each other. The spectral intensity of a spontaneous emission is averaged over xo. The averaged spectral intensity is shown to have only a very weakly expressed resonance structure.

1. **Introduction**

There are many articles devoted to undulators and strophotrons [1- 44] and references therein..
In the electric strophotron with scalar potential $\Phi(x) = \Phi_0(x/d)^2$, where $\Phi_0$ and $2d$ are the height and the width of the potential "trough" the spectral intensity of emission was found [4,45] in the most interesting case, along the $z$ axis. It has a form

$$\frac{dE_\omega}{d\omega do} = \frac{T^2 e^2 \omega^2}{16\pi^2}\left(\alpha^2 + x_0^2 \Omega^2\right)\sum_{s=0}^{\infty}\frac{\sin^2 u_s}{u_s^2}\left(J_s(Z) - J_s(Z)\right)^2 \tag{1}$$

where $do$ is an infinitely small solid angle in the direction $0z$, and T is the time it takes for the electron to travel through the strophotron, $x_0$ and $\dot{x}_0 \approx \alpha$ are the initial transversal coordinate and speed of the electron, and frequency of transverse oscillations $\Omega^2 = \dfrac{2e\Phi_0}{\varepsilon_z d^2}$, $J_{s,s+1}(Z)$ are the Bessel functions,

$$u_s = \frac{T}{4\gamma^2}\left[\omega\left(1 + \frac{\gamma^2}{2}\left(\alpha^2 + x_0^2 \Omega^2\right)\right) - 2\gamma^2 \Omega(2s+1)\right] \tag{2}$$

$$Z = \frac{\omega}{8\Omega}\left(\alpha^2 + x_0^2 \Omega^2\right). \tag{3}$$

Equation (1) describes the spectrum of emission consisting of a superposition of the spectral lines located at the odd harmonic $(2s+1)\omega_{res}$ ($s = 0, 1, 2, ...$) of the main resonance frequency

$$\omega_{res} = \frac{2\gamma^2\Omega}{1+\frac{\gamma^2}{2}(\alpha^2 + x_0^2\Omega^2)}. \qquad (4)$$

These lines are separated each from other by the term $2\omega_{res}$ and their width is found from the condition $u_s = 1$ to be on the order of $2\omega_{res}/\Omega T$. If the number $N_0 = \Omega T = (L/\lambda_0)$ of electron oscillations at thel ength of the strophotron $L$ is large erlough $N_0 \gg 1$, the spectral lines in'(1) for a single electron do not overlap. A relative intensity of emission at the harmonics with a number (2s + 1) is determined by the factor $(J_s(Z) - J_s(Z))^2$. If $\alpha\gamma \ll 1$ and $|x_0|\Omega\gamma \ll 1$, $Z = (2s+1)/4\gamma^2(\alpha^2 + x_0^2\Omega^2) \ll s$. In this case of rather small $\gamma$, $\alpha$, and $|x_0|$, only the main frequency $\omega_{res}$ (4) is emitted with an appreciable intensity. Emission at higher harmonics s $\gg$ 1 becomes possible only when $\alpha\gamma \gg 1$ or $|x_0|\Omega\gamma \gg 1$. But, of course, in this case, the main resonance frequency $\omega_{res}$ (4) falls down.

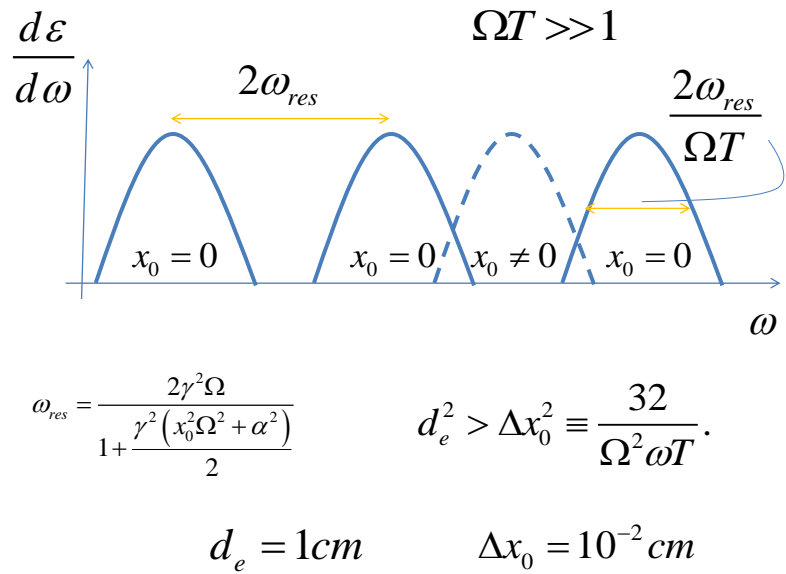

Fig.1. Dependence of spectral distribution of electron on frequency in relativistic strophotron.

Formally, these results completely coincide with the results of plane undulator [1,2,6]. Hence, for a single electron, the spectral intensity of a spontaneous emission can be reduced to the coinciding forms in the cases of the strophotron and of the plane undulator. But this coincidence will be shown to disappear when, instead of a single electron, one considers the electron beam as a whole. And the reason is in a different definition and parametrical dependence of the undulator and strophotron parameter in the undulator and in the strophotron.

## 2. Averaging over electron distribution

Although for a single electron, the spectral lines of emission (1) do not overlap as long as $\Omega T \gg 1$, they can overlap for different electrons, i.e., for different values of $x_0$. This possibility results from a strong dependence of $\omega_{res}$ (4) on $x_0$, which means that there is a strong inhomogeneous broadening of the spectral lines. To take this broadening into account, we must average (1) over $x_0$. This averaging will be ddone with the use of the Gaussian distribution function

$$f(x_0) = \frac{1}{\tilde{d}_e \sqrt{\pi}} e^{-x_0^2/\tilde{d}_e^2}, \qquad (5)$$

where $\tilde{d}_e = d_e / 2\sqrt{\ln 2}$, $d_e$ is t he diameter of the electron beam, $d_e < d$.

But first we will qualitatively estimate the conditions under which the inhomogeneous broadening exceeds the homogeneous width of the lines $2\omega_{res}/\Omega T$. Let, for example, the diameter $d_e$ not be too large and $x_0 = x_0^{(s)}$, Then the shift of $\omega_{res}$ (4) which occurs when $x_0$ grows from 0 to $d_e/2$ is equal t $\delta\omega_{res} = (1/16)d_e^2 \Omega \omega_{res}^2$. The corresponding shift of the line $(2s+1)\omega_{res}$ is $\delta\omega_s (1/16)d_e^2 \Omega \omega_{res}$. The shift $\delta\omega_s$ is larger than the homogeneous width $2\omega_{res}/\Omega T$ if

$$d_e^2 > \Delta x_0^2 \equiv \frac{32}{\Omega^2 \omega T}. \qquad (6)$$

If $d_e^2 > \Delta x_1^2 \equiv 32/\Omega\omega$, the shift $\delta\omega_s$ also exceeds the distance between' the adjacent lines $2\omega_{res}$ and the lines overlap. In the strophotron with external fields, the typical values are: $d_e = 1 cm$, $\Delta x_0 = 10^{-2} cm$, $\Delta x_1 = 3 \times 10^{-2} cm$ (see the Conclusion), and $d_e \gg \Delta x_0, \Delta x_1$ i.e., there is a strong inhomogeneous broadening and the lines overlap.

In the process of channeling in crystal, this effect can be less important because of a very small $d$ ($d = 10^{-8} cm$), e.g., for $\gamma = 10^3$ and $\omega = 2\gamma^2 \Omega$, $\Omega = 10^{14} s^{-1}$ and $\Delta x_1 = 3 \times 10^{-7} cm > d_e$. Broadening of spectral lines can be connected in this case with anharmonicity, but this effect is different from that resulting from the dependence $2\omega_{res}$ in (4). At lower energy, $\varepsilon = 10$-$50$ MeV, there is an even more pronounced difference between real and "macroscopic" channeling. For such energies, the number of levels in the potential pit ($e\Phi_0/\hbar\Omega$) in the case of a crystal is rather small ($\leq 10$ [10]). Hence, the anharmonicity is rather strong, and transitions between the levels can have an essentially quantum mechanical nature.

Vice versa, in the strophotron, the number of levels can be extremely large (up to $10^{12}$). In this case, the anharmonicity is very weak and well applicable.

Averaging (1) under the condition (6), we can replace the factor $\sin^2 u_s / u_s^2$ by $\delta u_s$ (some exclusions will be indicated below). From the condition $u_s = 0$, we find

$$\left(x_0^{(s)}\right)^2 = \frac{4}{\Omega\omega}(2s+1) - \frac{\alpha^2}{\Omega^2} - \frac{2}{\gamma^2\Omega^2} \equiv \frac{8}{\Omega\omega}(s - s_{\min}) \geq 0 \tag{7}$$

where

$$s_{\min} = \frac{\alpha^2\omega}{8\Omega} + \frac{\omega}{4\gamma^2\Omega^2} - \frac{1}{2}. \tag{8}$$

The argument of the Bessel functions $Z$ (3) is now equal to

$$Z\left(x_0 = x_0^{(s)}\right) \equiv Z_s = s + \frac{1}{2} - \frac{\omega}{4\gamma^2\Omega^2}. \tag{9}$$

To simplify the formulas, we will use the following assumptions

$$\alpha\gamma \gg 1, \quad \omega \gg 2\gamma^2\Omega \tag{10}$$

from which it follows, in particular, that $s_{\min} \gg 1$. Under these conditions, $Z_s < s$ (9) and both the index **s** of the Bessel function and its argument $Z_s$ are large $Z_s, s \gg 1$. Hence, the Bessel functions may be approximated by the McDonald functions [46]

$$J_s(Z_s) - J_{s+1}(Z_s) \approx \frac{2(s-Z_s)}{\pi\sqrt{3}\,s} K_{2/3}\left[\frac{2(s-Z_s)^{3/2}}{3\sqrt{s}}\right] \approx \frac{\omega K_{2/3}\left[\frac{\omega}{\omega_{\max}}\sqrt{\frac{s_{\min}}{s}}\right]}{2\gamma^2\Omega\pi\sqrt{3}} \tag{11}$$

*Where*

$$\omega_{\max} = 3\gamma^3\alpha\,\Omega \tag{12}$$

A similar approximation also has been used in the theory of a spontaneous emission in the plane undulator [2]. Now the averaged spectral intensity takes the form

$$\overline{\frac{dE_\omega}{d\omega do}} = \frac{e^2 T \omega^{5/2}}{2\sqrt{2}\, 3\pi^{3/2} \tilde{d}_e \Omega^{5/2} \gamma^4} \sum_{s-s_{min}\geq 1}^{\infty} \frac{s - s_{min} + \frac{\alpha^2 \omega}{8\Omega}}{s^2 \sqrt{s - s_{min}}}$$
$$\times K_{2/3}\left[\frac{\omega}{\omega_{max}}\sqrt{\frac{s_{min}}{s}}\right] \exp\left[-\frac{8(s - s_{min})}{\Omega \omega \tilde{d}_e^2}\right]. \tag{13}$$

If $\omega \ll \omega_{max}$ the McDonald function $K_{2/3}(u)$ may be approximated by

$$K_{2/3}(u) \approx \frac{\Gamma(2/3)}{2^{1/3} u^{2/3}}, \quad |u| \ll 1 \tag{14}$$

to give

$$\overline{\frac{dE\omega}{d\omega do}} = \frac{3^{1/3} \Gamma^2\left(\frac{2}{3}\right) e^2 T \left(\frac{\omega}{\Omega}\right)^{1/2}}{2^{1/6} \pi^{7/2} \tilde{d}_e}$$
$$\times \sum_{s-s_{min}\geq 1}^{\infty} \frac{s - s_{min} + \frac{\alpha^2 \omega}{8\Omega}}{s^{4/3} \sqrt{s - s_{min}}} \exp\left[-\frac{8(s - s_{min})}{\Omega \omega \tilde{d}_e^2}\right]. \tag{15}$$

All the terms of this series are well defined if $s_{min}$ is not close to an integer. In this case, the sum over s (15) may be replaced by the integral over the new variable $x = \left(8/\Omega\omega\tilde{d}_e^2\right)(s - s_{min})$ to yield

$$\overline{\frac{dE\omega}{d\omega do}} = \frac{3^{1/3} \Gamma^2\left(\frac{2}{3}\right) e^2 T \omega^{2/3}}{2^{2/3} \pi^{7/2} \alpha^{2/3}}$$
$$\times \int_0^{\infty} dx\, e^{-x} \frac{\frac{1}{\sqrt{x}} + \frac{\Omega^2 \tilde{d}_e^2}{\alpha^2}\sqrt{x}}{\left(1 + \frac{\Omega^2 \tilde{d}_e^2}{\alpha^2} x^2\right)^{4/3}}. \tag{16}$$

Rigorously, the lower limit of this integral is not zero $x_{min} = \left(\Delta x_0 / \tilde{d}_e\right)^2 \ll 1$ (what corresponds to the condition $s - s_{min} \geq 1$). But approximately $x_{min}$ may be replaced by 0. It gives an error $= \sqrt{x_{min}} \ll 1$ because the integrand of (16) only has a root singularity at the point x = 0 which is easily integrated.

Equation (11) describes the nonresonant background of a spontaneous emission of the electron beam in the strophotron. When $s_{min}$ is not close to an integer, the resonance structure of a spontaneous emission is completely smoothed out after averaging over $x_0$.

As a function of $\omega$, the spectral intensity (16) grows $\omega^{2/3}$ as long as $\omega < \omega_{max}$ When $\omega > \omega_{max}$ the McDonald function $K_{2/3}(u)$ in (13) becomes exponentially small for s not too far from $s_{min}$ This behavior of $K_{2/3}(u)$ cuts the sum over s from the region of small s by a condition stronger than $s > s_{min}$ Hence, in the integral over $x$ (16), the lower limit does not become small, $\tilde{x}_{min} = (\alpha / \tilde{d}_e \Omega)(\omega / \omega_{max})^2$.

If $\alpha \gg \tilde{d}_e \Omega$ the integral (16) itself becomes exponentially small. Thus, $\omega_{max}$ (12) is the frequency at which the spectral intensity $\overline{(d\mathrm{E}\omega / d\omega do)}_{n.r.}$, has its maximum which, for $\alpha = \tilde{d}_e \Omega$, is on the order of

$$\left[\left(\frac{\overline{d\mathrm{E}\omega}}{d\omega do}\right)_{n.r.}\right]_{max} \sim e^2 \Omega T \gamma^2. \tag{17}$$

Now let $s_{min}$ be close to some integer $|s_{min} - N| \ll 11$ where $N = [s_{min}]$ or $N = [s_{min}]+1$; [u] denotes the integer closest to u but less than u. In this case, the term with s = N in the sum (13) becomes anomalously large and tends to infinity when $s_{min} \to N$. This result indicates now that the factor $\sin^2 u_s / u_s^2$ in (1) may not be replaced by $\delta u_s$ because it has a singularity stronger than the $\delta$ function. This is the reason why the term with s = N can have a resonance character even after averaging over $x_0$. Taking separately only the resonance term s = N in the sum (1), we can average this term over $x_0$ directly:

$$\left(\frac{\overline{d\mathrm{E}_\omega}}{d\omega do}\right)_{res} = \frac{2\sqrt{2}e^2 T^{3/2} \omega_N^{5/2}}{3\pi^{9/2} \tilde{d}_e \Omega \alpha^2 \gamma^4} K_{2/3}^2 \left[\frac{\omega_N}{\omega_{max}}\right] F_{res}(\omega) \tag{18}$$

where, under the conditions (10), $\omega_N = (4\Omega / \alpha^2)(2N+1)$ and $F_{res}(\omega)$ is the resonance form factor

$$F_{res}(\omega) = \int d\xi \frac{\sin^2\left[\xi^2 + \frac{\alpha^2 T}{8}(\omega - \omega_N)\right]}{\left[\xi^2 + \frac{\alpha^2 T}{8}(\omega - \omega_N)\right]^2}. \tag{19}$$

Here we took into account that the interval $\Delta x_0$ giving the main contribution to the integral over $x$. is determined by (6) and according to the given assumption $\Delta x_0 \ll d_e$. For this reason, the exponential factor in the Gaussian distribution function $f(x_0)$ (5) was replaced by the unity.

When $\omega < \omega_{max}$ (18) with the aid of approximation (14) is reduced to

$$\left(\frac{\overline{dE_\omega}}{d\omega do}\right)_{res} = \frac{3^{1/3} 2^{5/6} \Gamma^2\left(\frac{2}{3}\right)}{\pi^{9/2} \tilde{d}_e} \frac{e^2 T^{3/2} \omega_N^{1/6} \Omega^{1/3}}{\alpha^{2/3}} F_{res}(\omega). \tag{20}$$

The resonances are located at $\omega \approx \omega_N = (2N+1)\omega_{res}$. In a general case, the given frequency of radiation $\omega$ can be realized by different electrons with different values of $x_0$. The larger $x_0$ is the smaller $\omega_{res}$ (4) is and the larger s are necessary to provide the given $\omega$. When $x_0 = 0$, $\omega_{res}$ is maximum and the necessary s is minimum. Hence, there is the threshold number s (equal to $N = [s_{min}]$ or $N = [s_{min}]+1$ ), and this threshold number is realized by the electrons with $x_0 = 0$. At the threshold, the intensity of each harmonic has its maximum because the factor $\sin^2 u_s / u_s^2$ has, in this case, a stronger singularity than the $\delta$ function. This anomalously large contribution of the *(2N + 1)* st harmonic at the threshold of its appearance means that even the averaged intensity of a spontaneous emission
concerns a resonance structure to some extent. The physical meaning of the parameter $\Delta x_0$ (6) is that not all the electrons contribute to the resonance part of emission, but only those which enter the strophotron near the axis ($x_0 = 0$) in a small interval of values of $x_0$, $x_0, |x_0| \leq \Delta x_0$. This is the reason why the height of these resonance peaks is not too large. From (16) and (20), we find

$$\left(\overline{dE}/d\omega do\right)_{res} / \left(\overline{dE}/d\omega do\right)_{n.r.} = \frac{T^{1/2}}{d_e \omega^{1/2}} \ll 1. \tag{21}$$

Hence, in a spontaneous emission, the nonresonant background is much higher than the resonance peaks when $N \geq N_0$.

The spectral width of the resonances is equal to the width of the curve $F_{res}(\omega)$ (19) $\Delta\omega = 8/\alpha^2 T = 2\omega_{res}(x_0 = 0)/\Omega T \ll 2\omega_{res}(x_0 = 0)$. This width is small. For this reason, remembering Madey's theorem [47], one can expect that in a stimulated emission, the role of resonances can be more important because the derivative from the narrow function $F_{res}(\omega)$ (19) can be large enough. This conclusion is confirmed by the following consideration.

## 3. Conclusion

The dependence of resonance frequency on electron transverse coordinate gives rise to a very strong inhomogeneous broadening of the spectral lines. The broadening can become large enough for the spectral lines to overlap with each other. The spectral intensity of a spontaneous emission is averaged over $x_0$. The averaged spectral intensity is shown to have only a very weakly expressed resonance structure.